\documentclass[a4paper,11pt,usenatbib]{article}
\pdfoutput=1 

\usepackage{jcappub} 

\usepackage[T1]{fontenc} 
\usepackage{lineno}
\usepackage{natbib}
\usepackage{hyperref}

\title{\boldmath Prospects for joint observations of gravitational waves and gamma rays from merging neutron star binaries}

\author[a,b]{B. Patricelli,}
\author[a,b]{M. Razzano,}
\author[b]{G. Cella,}
\author[a,b]{F. Fidecaro,}
\author[c]{E. Pian,}
\author[d,e]{M. Branchesi,}
\author[c,f]{A. Stamerra}

\affiliation[a]{Dipartimento di Fisica, Universit\`a di Pisa, \\Largo B. Pontecorvo, 3, 56127 Pisa, Italy}
\affiliation[b]{INFN - Sezione di Pisa, \\Largo B. Pontecorvo, 3, 56127 Pisa, Italy}
\affiliation[c]{Scuola Normale Superiore, \\Piazza dei Cavalieri, 7, 56126 Pisa, Italy}
\affiliation[d]{Universit\a di Urbino \\Via Aurelio Saffi, 2, 61029 Urbino, Italy}
\affiliation[e]{INFN - Sezione di Firenze, \\Via G. Sansone, 1, 50019 Sesto Fiorentino, Italy}
\affiliation[f]{INAF - Osservatorio Astronomico di Torino, \\Strada Osservatorio 20, 10025, Pino Torinese, Italy}

\emailAdd{barbara.patricelli@pi.infn.it}
\emailAdd{massimiliano.razzano@unipi.it}
\emailAdd{giancarlo.cella@pi.infn.it}
\emailAdd{francesco.fidecaro@unipi.it}
\emailAdd{elena.pian@sns.it}
\emailAdd{marica.branchesi@uniurb.it}
\emailAdd{stamerra@oato.inaf.it}

\abstract{
The detection of the events GW150914 and GW151226, both consistent with the merger of a binary black hole system (BBH), opened the era of gravitational wave (GW) astronomy. Besides BBHs, the most promising GW sources are the coalescences of binary systems formed by two neutron stars or a neutron star and a black hole. These mergers are thought to be connected with short Gamma Ray Bursts (GRBs), therefore combined observations of GW and electromagnetic (EM) signals could definitively probe this association. We present a detailed study on the expectations for joint GW and high-energy EM observations of coalescences of binary systems of neutron stars with Advanced Virgo and LIGO and with the \emph{Fermi} gamma-ray telescope. To this scope, we designed a dedicated Montecarlo simulation pipeline for the multimessenger emission and detection by GW and gamma-ray instruments, considering the evolution of the GW detector sensitivities. We show that the expected rate of joint detection is low during the Advanced Virgo and Advanced LIGO 2016-2017 run; however, as the interferometers approach their final design sensitivities, the rate will increase by $\sim$ a factor of ten. Future joint observations will help to constrain the association between short GRBs and binary systems and to solve the puzzle of the progenitors of GWs. Comparison of the joint detection rate with the ones predicted in this paper will help to constrain the geometry of the GRB jet.}


\begin{document}
\maketitle
\flushbottom


\section{Introduction}  

The recent detection of gravitational waves (GW) \citep{2016PhRvL.116f1102A,2016PhRvL.116x1103A} opened the era of GW astronomy. The two detected events, labeled GW150914 and GW151226, are both consistent with the coalescence of two black holes (BBH) at a distance of $\sim$ 400 Mpc.

During the fall of 2016 the Advanced Virgo detector \cite{2015CQGra..32g4001T} will also begin observations, and will start with LIGO a second joint observation run (``O2''). In the next years, both interferometers will be upgraded, and will progressively increase their sensitivity up to a factor of ten with respect to the initial LIGO and Virgo. Furthermore, also the KAGRA \cite{2013PhRvD..88d3007A} and the LIGO-India\footnote{https://www.ligo.caltech.edu/page/ligo-india} interferometers will become operative within the next six years, further expanding the frontiers of GW astronomy and the multimessenger investigation of cosmic sources.
Within this multimessenger context, the identification of an electromagnetic (EM) and/or a neutrino counterpart of a GW signal is fundamental to characterize the source and its progenitor. The detection of GW150914, and the subsequent possible gamma-ray counterpart found by $Fermi$-GBM \cite{2016ApJ...826L...6C} (but see also \cite{2016ApJ...827L..38G,2016ApJ...820L..36S}), opened a debate on the possibility of an EM counterpart for merging BBHs. 

Beside BBHs, the most promising GW sources expected to emit EM radiation are the coalescences of binary neutron stars (BNS) and black holes (NSBH). 
During the late stage of inspiraling, these systems are expected to emit GWs in the frequency range of ground-based interferometers ($\sim$ 10 Hz - 10 kHz). Furthermore, they are expected to be the most plausible candidate progenitors of short Gamma Ray Bursts (GRB) (e.g. \cite{1989Natur.340..126E,1992AIPC..272.1626P,1992ApJ...395L..83N,2012ApJ...746...48M}). Short GRBs are intense and highly variable flashes of $\gamma$ rays whose duration is $<$ 2 s (the $prompt$ emission), sometimes followed by a long lasting afterglow emission at lower energies. They are believed to be powered by ultra-relativistic jets produced by rapid accretion onto the central compact object formed during the coalescence. A coincident detection of a short GRB and a GW signal will be of paramount importance for several reasons. First, it will provide complementary information about the source: in fact, GW signal is the key to determine the mass distribution and the configuration of the gravitational field of the source, while the EM counterpart allows one to reconstruct the mechanisms for particle acceleration and emission, as well as to probe its environment. Furthermore, an EM counterpart will be crucial to pinpoint the host galaxy, thus determining the redshift and better constrain the parameters of the binary system.\\
There are two possible scenarios for joint GW and EM detections. In the scenario of the \emph{EM follow-up}, when a GW event is detected by the low-latency analysis pipelines, a prompt alert is issued to a network of EM observatories, that start observing the sky region consistent with the GW signal. The first broad-band EM follow-up campaign was performed to search for a counterpart of GW150914, but no firm counterpart was detected by any instruments \cite{2016arXiv160208492A}.
 
The second scenario is the \emph{externally-triggered} GW search, where a short GRB or another EM transient is detected, and GW data are analyzed in detail for a possible GW counterpart. 
The EM follow-up campaigns are challenging for several reasons. First, the sky localization provided by the ground-based interferometers is in order of hundreds of square degrees\footnote{The sky localization can be of the order of a few tens of square degrees for the most intense GW signals and is expected to be better with a higher signal to noise ratio; in particular, the major improvement in the sky localization will be obtained with the increase in the number of interferometers in the GW-detector network.} (see e.g. \cite{2014ApJ...795..105S,2016PhRvL.116f1102A}), therefore large FOVs are essential to properly cover the large GW error boxes. Furthermore, within such large GW error boxes, a huge number of optical transients is expected, making it difficult a clear and univocal association of an EM counterpart to the GW event (the number of optical transients spatially and temporally coincident with GW events is expected to be in order of hundreds, see e.g. \cite{2013ApJ...767..124N}). These difficulties are somewhat mitigated at gamma-ray energies, where large-FOV instruments like $Fermi$ are operative: the number of transient events at high energies is much smaller than at lower energies (for instance, in the \emph{Fermi}-GBM transient catalog there are only a few events in an area of 100 square degrees, see \cite{2016ApJ...826..228J}). 
 
Among the current $\gamma$-ray observatories in operation, \emph{Fermi} is one of those that better combines huge sky and energy coverage. \emph{Fermi} carries two instruments onboard: the Gamma-ray Burst Monitor (GBM, \cite{2009ApJ...702..791M}) and the Large Area Telescope (LAT, \cite{2009ApJ...697.1071A}). The GBM is specifically designed for GRB studies and can observe the full unocculted sky at energies between $\sim$8 keV and $\sim$40 MeV. The Large Area Telescope (LAT, \cite{2009ApJ...697.1071A}) is a pair-conversion telescope observing gamma rays of higher energies, from $\sim$ 20 MeV to more than 300 GeV. LAT has a large ($\sim$8000 cm$^{2}$) effective area and a wide (2.4 sr) FOV, large enough to cover a good portion of the GW error box even in the case only two interferometers record an event with high enough signal to noise ratio. Furthermore the sharp Point Spread Function (PSF) of the LAT (on-axis, 68\% containment radius at 10 GeV is $\sim$ 0.1$^\circ$) provides good localization of gamma-ray sources. Furthermore, if GBM detects a GRB above a fixed threshold\footnote{The on-board trigger threshold is $\sim$ 0.7 photons cm$^{-2}$ s$^{-1}$ \cite{2009ApJ...702..791M}.}, \emph{Fermi} slews to move the GRB into the FOV of the LAT. A detection by $Fermi$-LAT would provide a more precise localization of the GRB, thus allowing narrow-FOV optical telescopes to follow-up the event: in fact, the small \emph{Fermi}-LAT error region can be rapidly covered with a smaller number of exposures, in order to look for the putative galaxy host.


In the past many authors investigated the prospects for joint observations of GWs and short GRBs (see, e.g., \cite{2013ApJ...767..124N,2014MNRAS.437..649S,2015ApJ...799...69R,2015ApJ...809...53C}). A key ingredient for this study is represented by the merger rate of BNS systems in the local universe, that can be estimated with two complementary approaches: using accurate population synthesis modeling of the evolution of binary systems (see, e.g., \cite{2010CQGra..27q3001A} and references therein) or the local rate of short GRBs inferred from EM observations (see, e.g., \cite{2006A&A...453..823G,2006ApJ...650..281N,2014MNRAS.437..649S}).

On one side, BNS systems are strongly believed to be sources of GWs, and the comparison between predictions based on population synthesis models and the future GW detection rates will help to constrain the BNS merger rate, and thus to better understand the physics of these systems. Then, the comparison between the associated predicted EM detection rate with the observations could provide information about the short GRB progenitors: for instance, a smaller predicted EM detection rate with respect to the observed one could mean that BNS systems are not the only possible progenitors for short GRBs, and this could provide hints toward other interesting production channels; viceversa, a higher predicted rate could mean that not all the BNS can form a relativistic jet with a GRB emission. 

On the other hand, the comparison between predictions based on the local short GRB rate and the future joint EM and GW detections could help to constrain the jet opening angle of short GRBs ($\theta_j$). In fact, the EM emission from short GRBs is believed to be beamed and the observed sources are the on-axis ones, i.e. the ones for which the angle between the line-of-sight and the jet axis is less than $\theta_j$. Therefore, assuming that all the short GRBs have a BNS progenitor, the ratio between the GW and the EM detections strictly depends on the fraction of on-axis sources, and then on $\theta_j$: the comparison between the predicted and the observed rate of EM and GW detections could therefore be used to put constraints on its value.

In this work we present the prospects for joint GW and high-energy EM observations with Advanced Virgo, Advanced LIGO and \emph{Fermi} instruments, based on detailed simulations of BNS mergers accompanied by short GRBs; we used both the above described approaches to estimate the merger rate of BNS systems. This work differentiates from previous studies for several key aspects: first of all, this is the first time that joint EM and GW observations are investigated combining accurate population synthesis modeling (or accurate estimates of the local short GRB rates) with pipelines specifically developed to provide low-latency GW sky localization; furthermore, while previous works only focused on the GRB prompt emission, here we studied the possibility to detect the whole GRB emission (prompt and afterglow), with focus on gamma rays. The work is organized as follows. In Secs. \ref{sec:method} and \ref{sec:localRate} we explain our simulation and analysis pipeline, and in Sec. \ref{sec:results} we discuss the results and compare them with previous works. Finally, in Sec. \ref{sec:concl} we present our conclusion and further extension of the present work.

\section{Methods: simulating BNSs and their multimessenger detection}\label{sec:method}
In order to estimate the rates of joint high-energy EM and GW detections of merging BNS systems and investigate the high-energy follow-up scenarios, we designed a specific Montecarlo simulation pipeline for the BNS multimessenger emission and detection by GW and gamma-ray instruments. For our study, we focused on the case of GW detection by Advanced Virgo and Advanced LIGO, and EM detection by $Fermi$ instruments. This simulation pipeline is composed of three main steps: i) creation of a plausible ensemble of merging BNSs (Sec. \ref{sec:nsns}); ii) simulation of GW emission and detection by interferometers (Sec. \ref{sec:gw}); iii) simulation of associated short GRBs and detection with \emph{Fermi} (Sec. \ref{sec:GRB}). In order to estimate the detection rates, we simulated 1000 realizations, each one corresponding to an observing period of 1 year. The evaluation of the detection rates has been then repeated using different sensitivity thresholds of the ground-based interferometers, in the current stage and in the future, design stage.

\subsection{The merging BNS systems}\label{sec:nsns}
We generated a sample of synthetic galaxies populating the local universe accessible with Advanced Virgo and Advanced LIGO. In this work we only considered Milky Way-like galaxies and we did not include elliptical galaxies, expected to give a minor contribution to the BNS merger rate (e.g., \cite{2010ApJ...716..615O}). 
We used a constant galaxy density $\rho_{\rm gal}$=0.0116 Mpc$^{-3}$, extrapolated from the density of Milky Way equivalent galaxies in the local Universe \citep{2008ApJ...675.1459K}. Simulated galaxies have an isotropic and homogeneous distribution in space.  
Our galaxy sample extends to a maximum distance of 500 Mpc, consistent with the expected horizon for BNS mergers of Advanced Virgo and Advanced LIGO in their final configuration \citep{2016LRR....19....1A}. 
Each simulated galaxy was then populated with a sample of merging BNS systems. In the past years many investigations have been performed on the formation and evolution of binary systems of compact objects in different environments (i.e. in globular clusters, fields or young stellar clusters) using Monte Carlo simulations and/or population synthesis models (see, e.g., \cite{2003MNRAS.342.1169V,2012ApJ...759...52D,2013MNRAS.428.3618C,2014MNRAS.441.3703Z}). For our study, we used the public synthetic database Synthetic Universe\footnote{www.syntheticuniverse.org}, developed by \cite{2012ApJ...759...52D}. They investigated the evolution of binary systems that leads to the formation of binary systems of compact objects (BNS, NSBH and BBH) for a synthetic galaxy similar to the Milky Way (i.e. for a galaxy with the same age, star formation rate and stellar initial mass function of the Milky Way, see \citep{2012ApJ...759...52D} for details). 
Synthetic Universe is well suited for our study: for each binary system it provides an estimate of the merging time\footnote{The merging time is the sum of the time needed to form the two compact objects and the time for the two compact objects to coalesce.}, essential to know if the system contributes to the merger rate and several physical properties of the BNS systems such as, e.g., the mass, needed to simulate the expected GW signal (see Sec. \ref{sec:gw}); all these quantities have been consistently estimated through accurate modeling of the evolution of the systems. In particular,  
\cite{2012ApJ...759...52D} used different population synthesis models to account for the uncertainties in the physics of the binary evolution such as, for example, the common envelope phase and the wind mass-loss, and considered two metallicities: Z=Z$_{\odot}$ and Z=0.1 Z$_{\odot}$, where Z$_{\odot}$ is the solar metallicity. In this work we considered a stellar population composed by a 50\%- 50\% combination of systems with Z = Z$_{\odot}$ and Z = 0.1 Z$_{\odot}$, according to the bimodal distribution of the star formation in the last Gyr observed by the Sloan Digital Sky Survey \citep{2008MNRAS.391.1117P}. We chose as our reference model the so called ``standard model B'', that best estimates the key parameters of the physics of double compact objects\footnote{For instance, the ``standard model B'' uses the ``Nanjing'' estimate of the binding energy parameter $\lambda$ of the Common Envelope evolution \cite{2010ApJ...716..114X}.} \cite{2012ApJ...759...52D,2013ApJ...779...72D}. To take into account the uncertainties related to the physics of the systems, we also considered the models ``V12, A and B'' models (for systems with Z= Z$_{\odot}$) and the ``V2 A and V1 B'' models (for systems with Z=0.1 Z$_{\odot}$)\footnote{It is worth to mention that recently a new dataset, produced by \cite{2015ApJ...814...58D}, has been included in Synthetic Universe. Using the work done by \cite{2012ApJ...759...52D} as a reference, \cite{2015ApJ...814...58D} introduced more recent constraints on the initial conditions for young massive stars and investigated their impact on the binary merger rates. They found that the changes in the merger rates are negligible with respect to the evolutionary model uncertainties; furthermore, the new assumptions do not produce significant changes in the distributions of final component masses and merging time.}, where the merger rate are maximum and minimum. For each model and metallicity, if the systems merge within the age of the simulated host galaxy (assumed to be 10 Gyr \cite{2012ApJ...759...52D}), they are included in our sample. We then randomly extracted from the sample several BNS systems in accordance with the merger rates reported in \cite{2012ApJ...759...52D} and we populated the synthetic galaxies with them.

In \cite{2016ApJ...819..108B} it is shown that the merger rate density for BNS systems within the Advanced Virgo and Advanced LIGO range does not change in a significant way if the evolution of the star formation rate and of the metallicity through the cosmic time are taken into account. Therefore, for simplicity in this work we neglect the evolution of the merger rate with redshift. 


\subsection{The GW signals, their detections and sky localizations}\label{sec:gw}
To simulate the GW signals associated with the mergers we need: the mass, the sky position and the spin of the systems. We used the masses reported in \citep{2012ApJ...759...52D}. 
The coordinates and distances are the same as the host galaxy, and we gave each BNS a random inclination of the orbital plane with respect to the line of sight $\theta$. For simplicity, we considered non-spinning systems. 
This is a conservative approach: in fact, the introduction in our simulations of spinning BNS systems would result in smaller sky localization areas with respect to non-spinning systems, since the GW parameter estimation degeneracy is expected to be reduced (see, e.g., \cite{2009CQGra..26k4007R}). Furthermore, only low spin BNS have been observed up to date: the most rapidly rotating pulsar found in a binary system, i.e. PSR J0737-3039A, has a period of $\sim$ 22.7 ms \citep{2003Natur.426..531B,2012PhRvD..86h4017B}, corresponding to a very low spin: $\chi \sim$ 0.05\footnote{$\chi$ is  defined as $cJ/GM^2$, where $c$ is the speed of light, $G$  is the gravitational constant and $J$ and $M$ are the angular momentum and the mass of the star respectively.}; however, it is worth to mention that the fastest-spinning millisecond pulsar that has been observed, i.e. PSR J1748-2446ad, has a lower period of $\sim$ 1 ms \cite{2006Sci...311.1901H} and then a higher spin: $\chi \sim$ 0.4.

For each merging BNS system, we simulated the expected GW inspiral signals using the ``TaylorT4'' waveforms (see, e.g., \cite{2009PhRvD..80h4043B}), that are constructed using post-Newtonian models accurate to the 3.5 order in phase and 1.5 order in amplitude. Then, we added the GW signal to the detector noise. Berry et al. (2015) \cite{2015ApJ...804..114B} investigated the expected performances of the GW detection pipelines using both a realistic noise and an ideal Gaussian noise, finding negligible differences when a threshold in the signal-to-noise ratio of 12 is considered (see below). Therefore, for simplicity in this work we used a Gaussian noise, as in \cite{2014ApJ...795..105S}. 
We used the sensitivity curves of Advanced Virgo and Advanced LIGO \cite{2016LRR....19....1A}. In particular, we focused on two of the future configurations of the detectors: the 2016-2017 configuration\footnote{A six-month science run is expected to take place with this configuration (the so called ``O2'').} and the final design configuration, expected to be achieved in 2019 and 2021 by Advanced LIGO and Advanced Virgo respectively. For the 2016-2017 configuration we used the noise power spectral density (PSD) curves in the middle of the ranges reported in Fig. 1 of \cite{2016LRR....19....1A}; for the design configuration we used the noise PSD curves  reported in Fig. 1 of \cite{2016LRR....19....1A}. 

The data obtained in this way have been then analyzed with the matched filtering technique \citep{wainstein63}. With this technique the data from all detectors are Wiener-filtered  with a bank of modeled templates, constructed with different choices of the intrinsic parameters (e.g. the masses, the inclination angle etc) of the binary systems. The output is an estimate of the signal-to-noise ratio (SNR) $\rho$ with respect to that template in that detector:
\begin{equation}
\rho=\sqrt{4 \int_0^{f_{\rm ISCO}} \frac{|\tilde{h}(f)|^2}{S_n(f)}df},
\end{equation}
where $f_{\rm ISCO}$ is the frequency which corresponds to the innermost stable circular orbit of the system, $|\tilde{h}(f)|$ is the frequency-domain GW waveform amplitude and $S_n(f)$ is the noise PSD (see e.g. \cite{2010CQGra..27q3001A}). The signal is considered as a GW candidate if it produces a SNR above a given threshold in at least two detectors, 
with a time delay between the detectors consistent with the propagation of GWs.

We constructed templates specifically designed to detect our simulated signals, e.g. with the same intrinsic parameters used for the simulated signals. This choice is computationally less expensive than using the complete template banks, that cover a wide range of possible values of the intrinsic parameters, and does not affect in a significant way the results presented here. In fact, the template banks actually used in the analysis of GW data collected by Advanced Virgo and Advanced LIGO are constructed in such a way to cover the whole parameter space, with a spacing between the grid templates good enough so that the loss in the detection rate because of the discreteness of the bank is below $\sim$ 10 \% (see e.g. \cite{2012ApJ...748..136C} and references therein). 

We imposed a combined SNR threshold $\rho_c=\sqrt{\sum_i \rho_i^2}$=12, where $\rho_i$  is the SNR in the interferometer $i$; this corresponds to a false alarm rate (FAR) $<$ $10^{-2}$ yr$^{-1}$ \citep{2016LRR....19....1A}, which is affected in a negligible way by our simplified analysis without template banks. For each GW simulated candidate we then estimated the associated sky localization. Different algorithms have been developed by the Virgo collaboration and the LIGO scientific collaboration to calculate the sky maps of the GW events: cWB, LIB, LALInference and BAYESTAR \cite{2016arXiv160208492A}. Among them, the most accurate and sensitive for binary merger signals is LALInference. It constructs posterior probability distributions for the parameters of the binary system by matching the GW templates to the detector strain \cite{2015PhRvD..91d2003V}. However, it is not the best suited for the EM follow-up because of its latency, from hours to days. We chose to use 
BAYESTAR, that is a rapid Bayesian position reconstruction code that computes source location using the output from the detection pipelines; it is less accurate than LALInference, but it is fast and allows the prompt GW alerts to the EM telescopes \citep{2014ApJ...795..105S}. 

We considered two cases: an optimistic one, in which each interferometer is operating with a 100\% duty cycle (DC) and a more realistic case, in which each GW detector has an independent 80 \% DC\footnote{This is consistent with what was achieved with the initial detectors, where the two interferometers of LIGO had duty cycles of 78 \% and 67 \% in the S5 observing run \cite{2009RPPh...72g6901A}. Virgo had duty cycles of 81 \%, 80 \% and 73 \% in the observing runs VSR1-3 respectively \cite{2012CQGra..29o5002A}.} (see \cite{2016LRR....19....1A}).


\subsection{Simulated GRBs and electromagnetic detection with \emph{Fermi}}\label{sec:GRB}
In this work we assumed that all the BNS mergers are associated with a short GRB. Another possibility is that only a subset of them produce short GRBs (see e.g. \cite{2013ApJ...762L..18G}): the comparison of the detection rate estimated under our assumption with future GW and EM observations will help to constrain the jet opening angle ($\theta_j$)  and possibly the fraction of BNS that are actually progenitors of short GRBs.
We also assumed that the GRB jet is beamed perpendicular to the plane of the binary's orbit (i.e., that the angle of the observer with respect to the jet is equal to the inclination angle of the BNS system $\theta$, see e.g. \cite{2013MNRAS.430.2121P}). In the following we investigate separately the detectability of the GRB prompt and afterglow emissions with \emph{Fermi}. 

\subsubsection{The prompt emission}\label{sec:prompt}

GRB jets are characterized by initial Lorentz factors $\Gamma > 100$ (see e.g. \cite{2004RvMP...76.1143P}): therefore, because of the relativistic beaming, the prompt emission can be detected only if the GRBs are on-axis ($\theta \lesssim \theta_j$). We then assumed that the GRB prompt emission is constant within $\theta_j$, and zero outside. The value of $\theta_j$ is usually inferred from observation of a break in the afterglow light curve\footnote{When $\Gamma^{-1}=\theta_j$ a steepening in the flux decay of the afterglow emission is expected to be observed (the so called ``jet break'', see e.g. \cite{2004RvMP...76.1143P}).}; also, the lack of such a break is used to put lower limits on $\theta_j$. 
The lowest opening angle is the one inferred for GRB 090510, estimated to be between $\sim$ 0.1$^{\circ}$ and 1$^{\circ}$ \citep{2010ApJ...720.1008C,2010MNRAS.409..226K,2011MNRAS.414.1379P}. The highest lower limit for the opening angle is the one estimated for GRB 050724: this burst has no observed break after 22 days, leading to $\theta_j >  25^{\circ}$ \citep{2012MNRAS.425.2668C}. Finally, there are numerical studies suggesting that $\theta_j \leq 30^\circ$ (see, e.g., \cite{2011ApJ...732L...6R}). We therefore considered  $0.3^\circ \leq \theta_j \leq 30^{\circ}$ ($0.3^\circ$ is the value estimated for GRB 090510 by \cite{2011MNRAS.414.1379P}) and we adopt as our fiducial value $\theta_j=10^{\circ}$ (\cite{2014ApJ...780..118F,2015ApJ...813...64D}).

The duration of the short GRB prompt emission is much smaller ($<$ 2 s) than the minute-long latency needed to send a GW alert, so the prompt emission can be detected only if the GRB is already in the FOV of the detector (see e.g. \cite{2016ApJ...826L...6C,2016arXiv160208492A,2016ApJ...820L..36S}). \emph{Fermi}-GBM continuously observes the whole unocculted sky, so it is well suited to this purpose: to investigate the impact of the GBM sensitivity on the detectability of the simulated short GRBs, we estimated the lowest expected brightness of these sources. Following \cite{2015MNRAS.448.3026W}, we define the brightness as the 64-ms peak photon flux P$_{64}$\footnote{64 ms is the time interval over which counts from each detector of \emph{Fermi}-GBM are accumulated when a burst trigger occurs (CTIME data, see \cite{2009ApJ...702..791M}).} from the prompt emission in the 50-300 keV energy band. It is related to the 64-ms luminosity in the 1 keV-10 MeV energy band, $L$, by the following relation:
\begin{equation}
L=4\pi D(z)^2 (1+z) \frac{\int_{\rm{1keV}}^{\rm{10MeV}} E N(E) dE}{\int_{\rm{50keV (1+z)}}^{\rm{300keV (1+z)}}N(E) dE}  {\rm P_{64}},
\end{equation}
where $D(z)$ is the proper distance of the source at redshift $z$ and $N(E)$ its spectrum in the rest frame (see \cite{2015MNRAS.448.3026W}).
To estimate the lowest possible value of P$_{64}$, we considered a source located at our maximum distance of 500 Mpc and characterized by \mbox{$L=2.2  \times 10^{50}$ ergs}, the lowest luminosity of the short GRBs with known redshift (\cite{2015MNRAS.448.3026W}). We then assumed that $N(E)$ can be described by the Band function \citep{1993ApJ...413..281B}, with the parameters estimated for \emph{Fermi}-GBM bursts: E$_{\rm peak}$=800 keV (in the source frame), $\alpha_{\rm BAND}$=-05 and $\beta_{\rm BAND}$=-2.25 (see \cite{2011MNRAS.415.3153N,2015MNRAS.448.3026W}). With these assumptions we found P$_{64, \rm{min}} \sim$ 5 ph cm$^{-2}$ s$^{-1}$. This value is greater than the lowest P$_{64}$ measured for a short GRB by \emph{Fermi}-GBM, of 0.75$\pm$0.25 ph cm$^{-2}$ s$^{-1}$ (\cite{2012yCat..21990018P}): this means that \emph{Fermi}-GBM is sensitive enough to detect a short GRB with a P$_{64}$ as low as the lowest possible value of the GRBs in our sample.

Therefore, in the following we assume that all the GRBs would be observed by \emph{Fermi}-GBM if they were on-axis and in the FOV of the instrument (see also \cite{2015ApJ...809...53C,2016MNRAS.455.1522E}). To estimate the EM detection rates we took into account that GBM monitors the sky with a FOV of 9.5 sr and \mbox{DC $\sim$ 50 \%}\footnote{http://fermi.gsfc.nasa.gov/ssc/observations/types/grbs/}: this means that the fraction of on-axis GRBs that can be detected is \mbox{$\epsilon_{\rm FOV} \times$ DC}, where $\epsilon_{\rm FOV}$ is the FOV divided by $4 \pi$.

\subsubsection{The afterglow emission}\label{sec:afterglow}
We assumed that all the short GRBs have an afterglow emission at high energies (E $>$ 100 MeV). 
The afterglow emission is long lasting and can be potentially detected with observations triggered by GW alerts. We investigated the detectability of the high-energy afterglow emission with \emph{Fermi}-LAT: its FOV is large enough to cover the GW error box with a few tiled exposures and it has a better sky localization with respect to GBM. 

We simulated the GeV afterglow light curve and spectrum of GRBs using GRB 090510 as a template: in fact, this to date is the only short GRB to show emission up to GeV energies and, in particular, to show an extended emission ($\sim$ 200 s) at high energies\footnote{It is unclear whether the detection of only one short GRB with GeV afterglow emission is related to peculiar properties of the source (for instance, GRB 090510 is extremely energetic compared to other short GRBs: its isotropic energy is $\sim$ 10$^{53}$ erg \cite{2010ApJ...716.1178A}) and/or to observational issues. Therefore, the rates obtained under the assumption that all short GRBs have an extended high-energy emission, as GRB 090510, should be considered as upper limits.} ($\sim$ 4 GeV) \citep{2010ApJ...716.1178A}; the overall high-energy emission of GRB 090510 has been interpreted as afterglow emission (see, e.g., \cite{2010A&A...510L...7G}). After the peak, the GeV flux decays as a power law; the whole light curve can be fitted with a smoothly broken power law:
\begin{equation}
F(\rm{t})=A \frac{(\rm{t/t_{peak}})^{\alpha}}{1+(\rm{t/t_{peak}})^{\alpha+\delta}}.
\end{equation}
Fixing $\alpha=2$ according to the fireball model of GRBs \citep{1999ApJ...520..641S}, we found $A$=0.07$\pm$0.01 ph cm$^{-2}$ s$^{-1}$, $\delta=$1.60$\pm$0.15  and t$_{\rm peak}$=0.30$\pm$0.04 s (see also \cite{2010A&A...510L...7G}). 
The GeV spectrum is well described by a power law. In particular, the spectrum of the emission after the peak shows no significant evolution and it is well fitted by a power law with photon index $\beta$=-2.1 \citep{2010ApJ...709L.146D}. Before the peak the spectrum is harder but its spectral index is almost consistent, within the error, with -2.1 \citep{2010A&A...510L...7G}; therefore, for simplicity we assumed \mbox{$\beta$=-2.1} for the overall GeV emission. 

We simulated the GeV afterglow emission of the GRBs by correcting the observed light curve of GRB 090510 for the distance of the sources with respect to GRB 090510, whose redshift is z=0.903$\pm$0.001 \citep{2010A&A...516A..71M}. A further correction has been done to take into account that GRB 090510 is a uniquely bright burst: with a prompt emission isotropic energy $E_\gamma=3. 5\times 10^{52}$ ergs (excluding the LAT component, see e.g. \cite{2010A&A...510L...7G}), it is in fact among the most energetic GRBs ever observed. To do this correction, we simply re-scaled the GeV light curve for $E_\gamma$. In fact, assuming that the GeV emission is produced via synchrotron radiation above the cooling frequency (see e.g. \cite{2010A&A...510L...7G}), if the power-law index of the accelerated electrons is p=2 (see e.g. \cite{2001ApJ...561L.171P}) the observed flux $F$ is proportional to the kinetic energy $E_k$; since  $E_k \propto E_\gamma$ (see \cite{2001ApJ...560L.167P,2010A&A...510L...7G,2010MNRAS.403..926G}), it follows that $F \propto E_\gamma$. 
We considered the following values of isotropic energy: $10^{49}$ ergs, that is the minimum isotropic energy observed for a short GRB (see e.g. \cite{2015ApJ...815..102F}) 
 and 3.5$\times 10^{52}$ ergs, that is the value estimated for GRB 090510.

We assumed that the GeV afterglow emission has a duration of at least $\sim 10^3$ s. This is longer that the observed duration of the GeV extended emission of GRB 090510, but this could be due to the limited sensitivity of \emph{Fermi}-LAT; if GRB 090510 were occurred in the local universe, its flux could have been intense enough to be detectable for more than 200 s.

We assumed $\theta_j$=10$^\circ$ (see sec. \ref{sec:prompt}) and we didn't apply any further correction for the GRBs with $\theta  <\theta_j$, since in this case an observer should see a light curve very similar to that for an on-axis observer (\cite{2002ApJ...570L..61G}). The late afterglow emission could be observable also for sources with $\theta > \theta_j$: in fact, as the jet decelerates by sweeping up the interstellar medium, $\Gamma$ decreases, causing the visible region around the line of sight to increase with time (see e.g. \cite{1999ApJ...519L..17S}). However, the off-axis emission is expected to be weaker and to reach its maximum when $\Gamma \sim 1/\theta$ (see e.g. \cite{2002ApJ...570L..61G}): for $\theta > 10^\circ$, this happens on a time scale of days. Since in this work we focused on the low-latency EM follow-up, we did not consider the case of off-axis GRBs.

Once the simulated light curves are obtained, we investigated the possibility of detection with \emph{Fermi}-LAT. To do this, we estimated the total time $t_{\rm f}$ each GRB should be observed so that its fluence reaches the high-energy LAT sensitivity. We focused on the sensitivity (in the energy range 0.1-300 GeV) corresponding to a GRB localization at 1-$\sigma$ of 1 deg: this localization accuracy is good enough to allow for the EM follow-up of the event with other telescopes. For example, an error region of a few square degrees can be covered in a single or a few exposures with optical large FOV telescopes (such as La Silla QUEST \cite{2007PASP..119.1278B}, PTF \cite{2009PASP..121.1395L}, Pan-STARRS \cite{2002SPIE.4836..154K} and VST \cite{2002SPIE.4836...43C}) and air Cherenkov telescopes such as MAGIC, H.E.S.S. and VERITAS (\cite{2012APh....35..435A}), as well as with a few tens of tiled observations by the \emph{Swift} XRT telescope (\cite{2005SSRv..120..165B}).

To estimate the LAT sensitivity to GRBs the instrument response functions are needed. In this work we used the sensitivity estimated  with the ``Pass 7'' reprocessed instrument response function\footnote{http://www.slac.stanford.edu/exp/glast/groups/canda/archive/p7rep$\_$v15/lat$\_$Performance.htm. Recently, the \emph{Fermi}-LAT collaboration has completed the development of the ``Pass 8'' event-level analysis (see e.g. \cite{2013arXiv1303.3514A}) that, among various performance improvements, provides a better modeling of the instrument's energy response function; however, the LAT sensitivity to GRBs with this new function is not publicly available yet.}. This sensitivity has been obtained in the energy range 10-1000 keV by assuming that the source spectrum is a Band function with different possible values of the high-energy spectral index $\beta_{\rm BAND}$, ranging from -2.75 to -2.0, and for two hypothetical values of beaming angle: 0$^{\circ}$ (on-axis source) and 60$^{\circ}$. We focused on the sensitivity estimated for $\beta_{\rm BAND}$=-2.0, that is the value closer to the value of $\beta$ estimated for GRB 090510 and we extrapolated the corresponding sensitivity to the energy range 0.1-300 GeV. Since we considered only on-axis sources, we compared the fluence of the simulated GRBs with the extrapolated sensitivity for $\theta = 0^{\circ}$.
 
We considered different possible scenarios: i) the LAT working in the survey mode is already covering the sky region of the event; ii)  the LAT is working in the survey mode and the source enters its FOV some time after the GW trigger and iii) the GW source is outside the FOV of the LAT and a a re-pointing of the instrument is performed. For case i) we can assume that the EM and GW observations of the event start at the same time (therefore the latency is 0 s). For case iii) we assumed a conservative latency of 10 minutes\footnote{This latency comprises the $\sim$ 3 minutes needed by GW pipelines to generate the GW triggers (see \cite{2016PhRvL.116f1102A}) and a few additional minutes to generate the sky map and for the validation checks.}. This could apply also for case ii), although higher latencies are possible (see e.g. \cite{2016arXiv160208492A}). 

\subsection{Triggered GW searches based on EM detections}\label{sec:EMTrigg}

The rate of joint EM and GW detections can be potentially increased with GW searches externally triggered by EM detections. In fact, external EM triggers can help rule out false alarms and thus lowers the necessary SNR threshold for a GW detection (see, for instance, \cite{1993ApJ...417L..17K,2013PhRvD..87l3004K}), since they decrease the time window and sky area in which GWs need to be searched for. Specifically, if $\rho_c$ is the SNR threshold associated with a given FAR for untriggered GW searches, the SNR threshold for EM triggered GW searches ($\rho_c^{\rm trig}$) required to achieve the same FAR is:
\begin{equation}
\rho_c^{\rm trig}\sim \sqrt{2 \times \log \left[\exp\left(\frac{\rho_c^2}{2}\right) \frac{t_{\rm obs} \times \Omega}{t_{\rm obs,0} \times \Omega_0}\right]},  
\end{equation}
(see e.g. \cite{2015PhRvL.115w1101B}), where $\Omega_0$, $t_{\rm obs,0}$ and $\Omega$, $t_{\rm obs}$ are the sky region and the observation duration for untriggered and EM triggered GW searches respectively. We considered $\Omega_0 \sim$ 40000 deg$^2$ (all sky searches) and $t_{\rm obs,0}$=1 year. For EM triggered searches, the fraction of the sky to be analyzed is limited by the spatial resolution of the GW detectors, so we use $\Omega$=100 deg$^2$ (see, e.g., \cite{2013PhRvD..87l3004K}). Finally, we considered  $t_{\rm obs}=\delta t \times N_{\rm GRB}$, where $\delta t$ is the GW search time window around the EM GRB trigger and $N_{\rm GRB}$ is the number of short GRBs expected to be detected in the observation period (1 year). Here we used \mbox{$\delta t$= 6 s} (5 s prior to the GRB to 1 s after the EM trigger), which is wide enough to allow for uncertainties in the emission model and in the arrival time of the electromagnetic signal (see \cite{2012ApJ...760...12A}) and $N_{\rm GRB}\sim$1, that is within the range of the expected yearly rate of short GRB detection with \emph{Fermi}-GBM (see Sec. \ref{sec:EMresults-prompt}). 
With these values (and considering $\rho_c$=12, see Sec. \ref{sec:gw}) we obtained $\rho_c^{\rm trig} \sim$ 10, i.e. there is a reduction of the SNR threshold of about $\sim$ 17 $\%$. We estimated the rate of joint EM and GW detections with this value of SNR threshold.

\section{Alternative approach - from the local short GRB rate to the BNS rate}\label{sec:localRate}
We estimated the rate of high-energy EM and GW detections of BNS systems also with a different approach, that differs from the previous one in the step (i): the rate of BNS merging systems (${\rm R_{BNS}}$) in the local universe has been directly estimated from the local short GRB rate (${\rm \rho_{sGRB}}$), without using population synthesis models.
 
We assumed that all the short GRBs have a BNS progenitor (as done previously) and that ${\rm \rho_{sGRB}}$ is constant within the local universe; therefore, the rate of observed short GRBs can be related to the all sky rate of binary mergers through:
\begin{equation}
{\rm R_{BNS}}=\frac{{\rm \rho_{sGRB}}}{f_b} \times \frac{4 \pi D^3}{3},
\end{equation}
where $f_b=1-cos(\theta_j)$ is the beaming factor, that represents the fraction of GRBs that are on-axis and D is the maximum distance considered (D=500 Mpc, see sec. \ref{sec:nsns}).
 
In the past years there have been numerous efforts to estimate ${\rm \rho_{sGRB}}$ based on the luminosity function and the redshift distribution of short GRBs, as inferred from EM observations (see, e.g., \cite{2006A&A...453..823G,2006ApJ...650..281N,2009A&A...498..329G,2012MNRAS.425.2668C,2014MNRAS.437..649S,2015ApJ...809...53C,2015MNRAS.448.3026W,2016arXiv160707875G}). However, there are still large uncertainties in the value of ${\rm \rho_{sGRB}}$: the current estimates ranges from $\sim$ 0.1 Gpc$^{-3}$ yr$^{-1}$ (see, e.g., \cite{2006A&A...453..823G}) to 40 Gpc$^{-3}$ yr$^{-1}$ (see, e.g., \cite{2006ApJ...650..281N}). In this work we used three among the most recent estimates of ${\rm \rho_{sGRB}}$: the ones reported by \cite{2015MNRAS.448.3026W,2016arXiv160707875G}, obtained with updated samples of GRBs observed by CGRO BATSE, \emph{Swift} and \emph{Fermi}. Specifically, \cite{2015MNRAS.448.3026W} found ${\rm \rho_{sGRB}} \propto L_{\rm min}^{-0.95}$, where  $L_{\rm min}$ is the low-end cut-off of the luminosity function of short GRBs;  in this work we assumed $L_{\rm min}=2.2 \times 10^{50}$ ergs/s (see sec. \ref{sec:prompt}): this corresponds to ${\rm \rho_{sGRB}} \sim$ 1 Gpc$^{-3}$ yr$^{-1}$. \cite{2016arXiv160707875G} estimated ${\rm \rho_{sGRB}}$ with different approaches, that mainly differ in the assumption of correlation (no correlation) between the peak energy and the isotropic energy and luminosity of short GRBs. Here we considered the models that predict the minimum and maximum values of  ${\rm \rho_{sGRB}}$: the model ``a'' (${\rm \rho_{sGRB}}$= 0.2  Gpc$^{-3}$ yr$^{-1}$) and ``c'' (${\rm \rho_{sGRB}}$= 0.8 Gpc$^{-3}$ yr$^{-1}$). For each value of ${\rm \rho_{sGRB}}$, we estimated ${\rm R_{BNS}}$ for $\theta_j$ in the range $0.3^{\circ} \leq \theta_j \leq 30^{\circ}$, considering as a fiducial value $\theta_j=10^{\circ}$ (see sec. \ref{sec:prompt}).

To calculate the associated rates of GW and EM detections we used the same procedures discussed in Secs. \ref{sec:gw} and \ref{sec:prompt}. In particular, also in this case we can assume that \emph{Fermi}-GBM is sensitive enough to detect all the local short GRBs that are in its FOV\footnote{All the GRBs contributing to ${\rm \rho_{sGRB}}$ have a $\rm{P_{64}}$ greater than the lowest value measured by \emph{Fermi}-GBM. In fact, we used the estimate of ${\rm \rho_{sGRB}}$ obtained by \cite{2015MNRAS.448.3026W} for $L_{\rm min}=2.2 \times 10^{50}$ ergs/s, that corresponds to $\rm P_{64, min} \sim$ 5 ph cm$^{-2}$ s$^{-1}$; furthermore, the values of ${\rm \rho_{sGRB}}$ reported in  \cite{2016arXiv160707875G} have been estimated with a sample of \emph{Fermi} GRB with $\rm P_{64} > $ 5 ph cm$^{-2}$ s$^{-1}$ (see also sec. \ref{sec:prompt}).}; therefore, the rate of local short GRB detectable by \emph{Fermi}/GBM has been estimated as \mbox{${\rm R_{EM}}=\left({\rm \rho_{sGRB}} \times \frac{4 \pi D^3}{3} \right) \times \frac{FOV}{4 \pi} \times DC$} (see also the discussion in sec. \ref{sec:prompt}).


\section{Results}\label{sec:results}
We present our expectations for joint GW and EM detections of BNS mergers. Specifically, in Secs. \ref{sec:GWresults} we present the GW detection rate and sky localization of merging BNS systems for the 2016-2017 and the design configuration of Advanced Virgo and Advanced LIGO, while in Secs. \ref{sec:EMresults-prompt} and \ref{sec:EMresults-afterglow} we show the expected rates of EM and joint EM and GW detections considering the prompt and the afterglow emission of the short GRBs associated with the BNS mergers; these results have been obtained with the method explained in Sec. \ref{sec:method}. The results obtained using the local rate of short GRBs (Sec. \ref{sec:localRate}) are shown in Sec. \ref{sec:EMresults-localRate}.


\subsection{GW detections and sky localizations}\label{sec:GWresults}

The expected number of GW detections over the six-month science run planned for 2016-2017 and a 1-year science run with the design configuration, considering an independent 80 \% DC of the interferometers, are in the ranges between 0.001-0.7 and 0.04-15 respectively. These ranges reflect the uncertainty in the merger rate of BNS systems\footnote{The  merger rate of BNS systems is the dominant source of uncertainty in the estimated detection rates.} (see Sec. \ref{sec:nsns}). The values we found are consistent with the ones previously presented by \cite{2016LRR....19....1A} (see Tab. \ref{tab:GW}); however, it can be noted that the lower and upper limits of our ranges are smaller than the ones reported by \cite{2016LRR....19....1A}, as well as than the value reported by \cite{2014ApJ...795..105S}. This reflects the different range of BNS merger rate considered: in fact, \cite{2016LRR....19....1A} and \cite{2014ApJ...795..105S} assumed ($10^{-8} - 10^{-5}$) Mpc$^{-3}$ yr$^{-1}$ and $10^{-6}$ Mpc$^{-3}$ yr$^{-1}$ respectively, while in this work we use the more conservative range\footnote{The merger rate densities have been estimated as $\rho_{\rm gal}\times \left(R_{Z_{\odot}}+R_{0.1 Z_{\odot}}\right)/2$, where $R_{Z_{\odot}}$ and $R_{0.1 Z_{\odot}}$ are the BNS merger rate (expressed in yr$^{-1}$) at solar and sub-solar metallicity respectively, for the different theoretical models by \cite{2012ApJ...759...52D} used in this work (see Sec. \ref{sec:nsns}).} $2 \times 10^{-9} - 8 \times 10^{-7}$ Mpc$^{-3}$ yr$^{-1}$.
 
\begin{figure}[tbp]
\begin{center}
\includegraphics[scale=0.5]{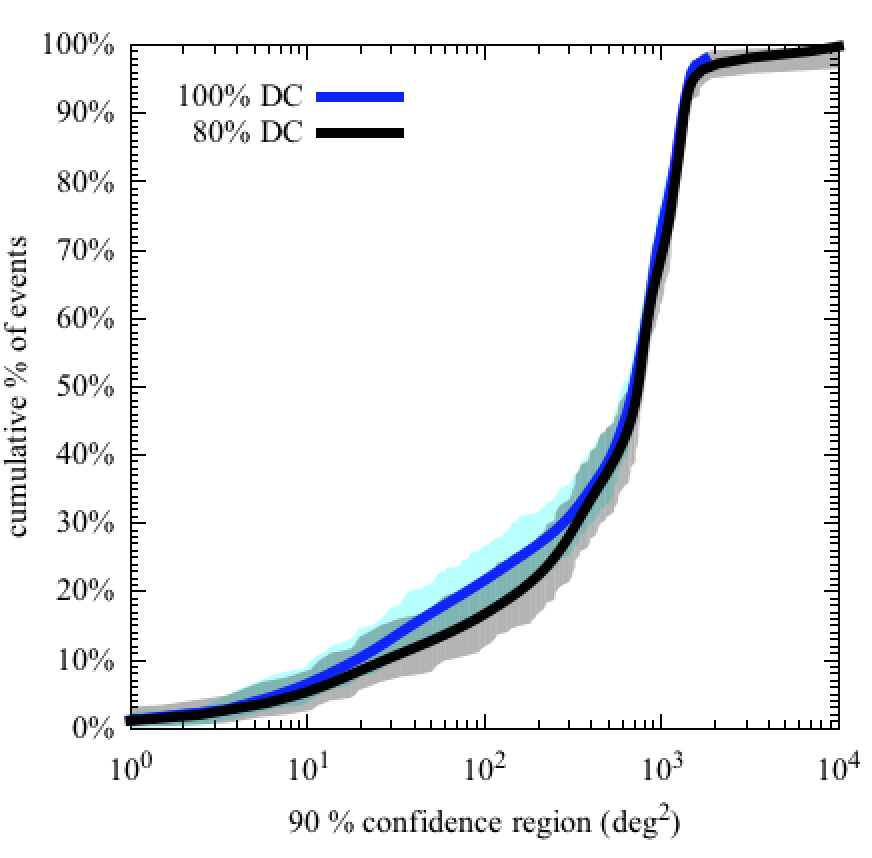}\includegraphics[scale=0.5]{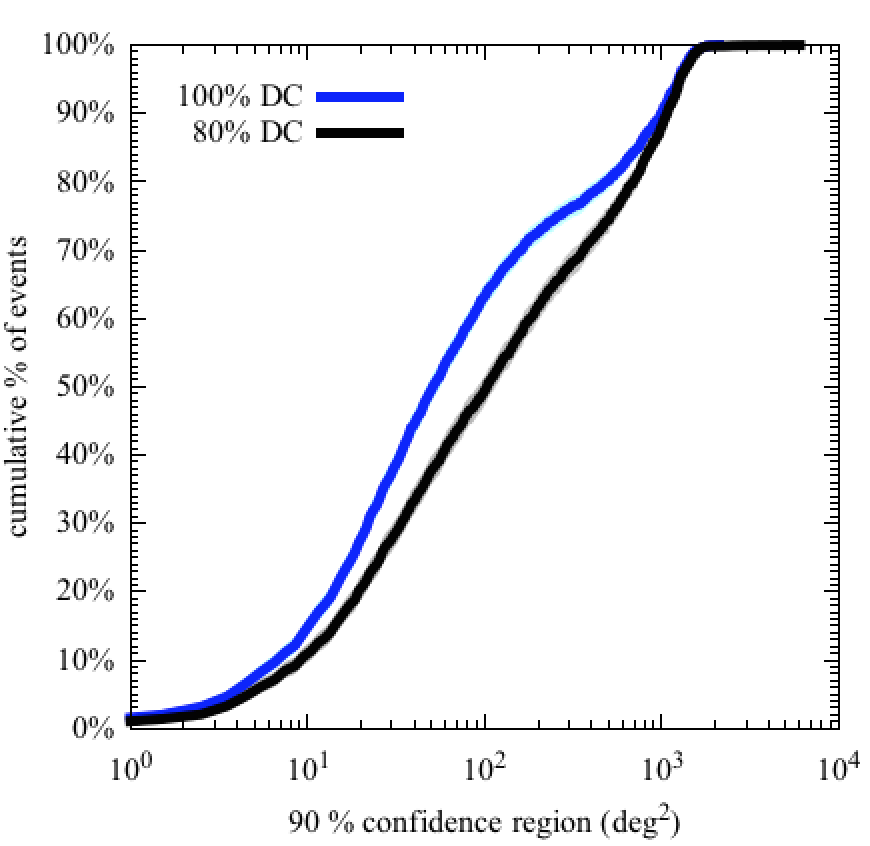}
\caption{\label{fig:A90} Cumulative histograms of sky localization areas of the 90 \% confidence region in the 2016-2017 (left) and in the design (right) scenarios, for a 100\% (blue) and an 80\% (black) DC. The shadowed regions enclose the 95 \% confidence intervals accounting for sampling errors, as computed from the quantiles of a beta distribution (see \cite{2011PASA...28..128C}). The Standard model B of \protect\cite{2012ApJ...759...52D} and a 50\%- 50\% combination of systems with Z= Z$_{\odot}$ and Z= 0.1 Z$_{\odot}$ have been considered.}
\end{center}
\end{figure}

To give an estimate of the accuracy of the sky localization of the GW candidates, that is critical for the EM follow-up, we calculated the expected cumulative number of detections as a function of the areas, in deg$^2$, inside of the 90 $\%$ confidence regions. These contours were constructed with the ``water-filling'' algorithm introduced in \cite{2014ApJ...795..105S}: we sampled the sky maps using equal-area pixelization \cite{2005ApJ...622..759G}, 
then we ranked these pixels from the most probable to least, and finally we counted how many pixels summed to the chosen probability. The results are shown in Fig. \ref{fig:A90} and summarized in Tab. \ref{tab:GW}, together with a comparison with other estimates reported in literature. 

It can be seen that the percentage of GW candidates with a good sky localization ($\lesssim$ 5$^\circ$, comparable with the accuracy of \emph{Fermi}-GBM) is expected to be only of 3 \% and 5 \% for the 2016-2017 and the design configuration respectively, consistently with the results reported by \cite{2016LRR....19....1A,2014ApJ...795..105S}; this corresponds to 0.001 and 0.05 events per year\footnote{These values refer to the standard model B of \cite{2012ApJ...759...52D}.}. The majority of the GW events are then expected to have a sky localization of the order of hundreds to thousands of square degrees (see Tab. \ref{tab:GW}). 

From Fig. \ref{fig:A90} it can also be noted that, as expected, the sky localization is better when a DC=100\% is assumed. This is mainly related to the fact that the greater is the number of GW interferometers detecting the event, the better is the sky localization (see e.g. \cite{2016LRR....19....1A}). When a DC=100\% is assumed, the three detectors are continuously working all the time, so there is a higher probability for a sufficiently intense GW signal to be detected by all of them. By contrast, when DC=80\%, there are intervals of time when only two GW interferometers are working (corresponding to $\sim 13 \%$ of the time for each pair of interferometers) and GW signals occurring in these time windows could only trigger two detectors. The improvement with DC is significant when the design configuration is considered. Specifically, the increase in the expected percentage of GW events detected by 3 interferometers is of $\sim$ 40 \%; consequently, also the sky localization is significantly better: for instance, the percentage of events having a sky localization $\leq$ 50 deg$^2$ increases by $\sim$ a factor of 3. This underline the importance of keeping the GW detectors in operation with continuity, in order to maximize their DC. The improvement is less significant when the 2016-2017 configuration is considered: in this case, the increase in the expected percentage of GW events detected by 3 interferometers is only of $\sim$ 20 \%. This is because of the fact that the 2016-2017 sensitivity of Advanced Virgo is not high enough to detect the less intense events, that are therefore expected to trigger only the two detectors of Advanced LIGO.

\begin{table}[tbp]
\begin{center}
\scalebox{0.88}{
\begin{tabular}{lllccccccccc}
\hline \hline   \vspace{0.1mm}\\
Configurations & Work & Number of BNS  & \% of BNS   & \% of BNS  & \% of BNS   & \% of BNS  \vspace{1mm}\\  
 & &  detections & with Loc. & with Loc. & with Loc. & with Loc.\vspace{1mm}\\  
  & &  (yr$^{-1}$)& $\leq$ 5 deg$^2$ & $\leq$ 20 deg$^2$ & $\leq$ 100 deg$^2$ & $\leq$ 1000 deg$^2$ \vspace{1mm}\\ 
\hline
\vspace{1mm}\\   
 & This work  & 0.05 (0.001 - 0.7) & 3 & 9 & 16 & 70\vspace{1mm}\\ 
2016-2017 &   \cite{2014ApJ...795..105S}\footnote{These estimates refer to the 2016 scenario.} & 1.5 & 2 & 8  &15 & - \vspace{1mm}\\  
  & \cite{2016LRR....19....1A} & 0.006-20 &2 &14  & - & - \vspace{1mm}\\  
 
\hline
\vspace{1mm}\\  
 2019+ (design) & This work  & 1 (0.04 - 15)  & 5  &  21   &50 & 90\vspace{1.1mm}\\  
 &  \cite{2016LRR....19....1A}& 0.2-200 &$>$ 3-8 & $>$ 8-30   & - & -\vspace{1mm}\\  
 
\hline\hline
\end{tabular}
}
\caption{\label{tab:GW} Expected GW detection rate and source localization for the 2016-2017 and the 2019+ (design) configurations, with an independent 80\% duty cycle of each interferometer, as assumed in \protect\cite{2016LRR....19....1A} and \protect\cite{2014ApJ...795..105S}. For the 2016-2017 configuration, our estimated number of BNS detections has been re-scaled to a 6-months observation period, to do a direct comparison with \protect\cite{2016LRR....19....1A} and \protect\cite{2014ApJ...795..105S}. The reported values refer to the Standard model B, for a 50\%- 50\% combination of systems with Z= Z$_{\odot}$ and Z= 0.1 Z$_{\odot}$; the range of GW detection rates reported in parenthesis has been estimated considering the range of BNS merger rates reported by \protect\cite{2012ApJ...759...52D} (see Sec. \ref{sec:gw}).} 


\end{center}
\end{table} 

\subsection{Joint EM and GW detections}

\subsubsection{GRB prompt emission}\label{sec:EMresults-prompt}

The rates of EM detections of GRB prompt emission with \emph{Fermi}-GBM and the rates of joint EM and GW detections are reported in Table \ref{tab:EM-GBM}, for different values of $\theta_j$. It can be seen that the expected number of joint EM and GW detections is less than one for the 2016-2017 scenario; however, when the final design configuration will be reached by Advanced Virgo and Advanced LIGO, there will be a greater chance of joint EM and GW detections for a 1 year science run (for instance, for $\theta_j$=30$^\circ$ the number of joint detections is expected to be between 0.003 and 2.6). The rates are sensitive to the value of the jet opening angle and, as expected, they increase when a larger value of $\theta_j$ is assumed. 


\begin{table}[tbp]    
\begin{center}
\scalebox{0.88}{
\begin{tabular}{lllcccccc}
\hline \hline   \vspace{0.1mm}\\
$\theta_j$ & EM & EM and GW & EM and GW \vspace{1mm}\\  
 & & \textbf{2016-2017} & \textbf{design}\vspace{1mm}\\ 
deg&  yr$^{-1}$ &  yr$^{-1}$ & yr$^{-1}$ \vspace{1mm}\\  
\hline
\vspace{1mm}\\  
 0.3 & $< 10^{-3}$ ($< 10^{-3}$) &   $< 10^{-3}$ ($< 10^{-3}$)  & $< 10^{-3}$ ($< 10^{-3}$)  \vspace{1mm}\\
 & $< 10^{-3}$ - 0.006 ($< 10^{-3}$ - 0.002)&  ($< 10^{-3}$ - $< 10^{-3}$) &  ($< 10^{-3}$ - $< 10^{-3}$) \vspace{1mm}\\
 \hline \vspace{0.1mm}\\
10 & 0.5 (0.2) & 0.002 (0.001) & 0.06 (0.03)\vspace{1mm}\\
  & 0.02 - 6.6 (0.003 - 2.4) & $< 10^{-3}$ - 0.04 ($< 10^{-3}$ - 0.02) & 0.002 - 0.9 ($< 10^{-3}$ - 0.5)  \vspace{1mm}\\
  \hline \vspace{0.1mm}\\
 30  & 4 (1.5) & 0.02 (0.007)  & 0.6 (0.2) \vspace{1mm}\\
  & 0.1 - 59 (0.02 - 22)& $< 10^{-3}$ - 0.4 ($< 10^{-3}$ - 0.1) & 0.02 - 7.6 (0.003 - 2.6) \vspace{1mm}\\  
\hline\hline
\end{tabular}
}
\caption{\label{tab:EM-GBM} Expected rates of EM and GW detections, considering a 80 $\%$ duty cycle of the interferometers and a 4$\pi$ (9.5 sr) FOV and 100 $\%$ (50 $\%$) duty cycle of \emph{Fermi}-GBM. Only the GRB prompt emission has been considered. The reported values refer to the Standard model B, for a 50\%- 50\% combination of systems with Z= Z$_{\odot}$ and Z= 0.1 Z$_{\odot}$. The range of rates reported for each value of $\theta_j$ have been estimated considering the range of BNS merger rates reported by \protect\cite{2012ApJ...759...52D} (see Sec. \ref{sec:gw}).} 

\end{center}
\end{table}

\begin{figure}[tbp]
\begin{center}
\includegraphics[scale=0.9]{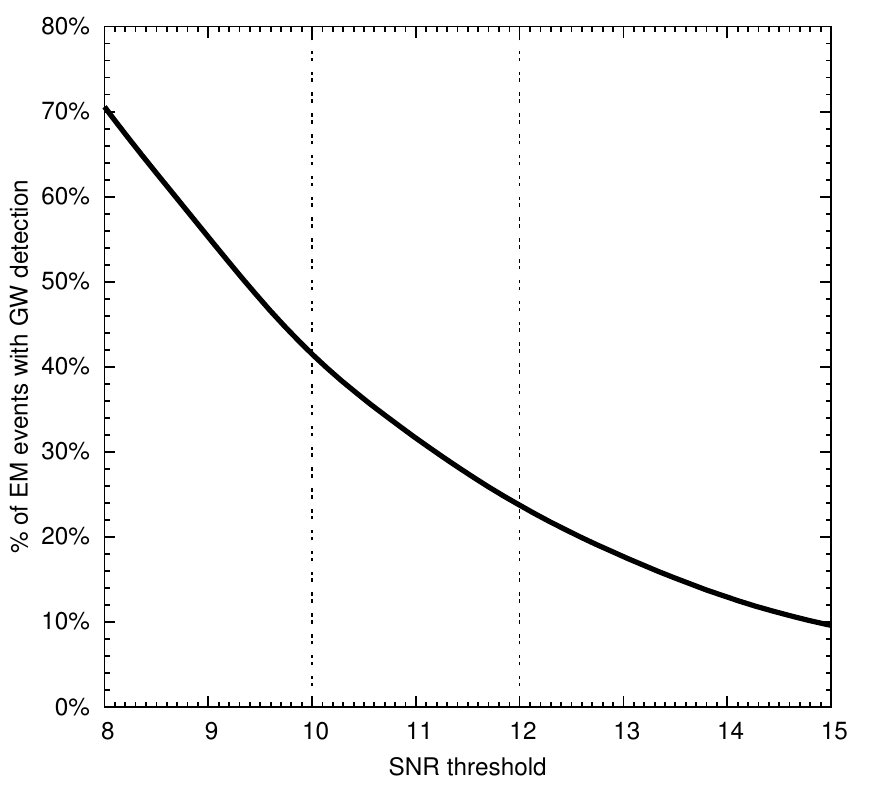}
\caption{\label{fig:SNR} Percentage of short GRB occurring within a maximum distance of 500 Mpc detectable by \emph{Fermi}-GBM that also have an associated GW detection, for different SNR threshold. The Standard model B of \protect\cite{2012ApJ...759...52D}, a 50\%- 50\% combination of systems with Z= Z$_{\odot}$ and Z= 0.1 Z$_{\odot}$, $\theta_j$=10$^{\circ}$, a 9.5 sr FOV and 50 $\%$ duty cycle of \emph{Fermi} and the design configuration of Advanced Virgo and Advanced LIGO have been considered. The vertical dashed lines mark the SNR threshold of 10 and 12 (see Sec. \ref{sec:EMTrigg}).}
\end{center}

\end{figure}

We compare our results with other recently published works. \cite{2015MNRAS.448.3026W} estimated the rate of joint EM and GW detections of short GRBs with \emph{Fermi}-GBM and Advanced Virgo and Advanced LIGO, considering events within 300 Mpc and assuming different values for the minimum GRB peak luminosity $L$. For \mbox{$L=2.2 \times 10^{50}$} ergs (the same used in this work, see Sec. \ref{sec:prompt}), they reported a rate of 0.11 $\pm$ 0.04 yr$^{-1}$. To do a comparison with this work, we re-calculated our rate of joint EM and GW detections for the same maximum distance; for $\theta_j=10^\circ$ and the design configuration we found the range of values ($< 10^{-3}$-0.3) yr$^{-1}$, which  is consistent with the value found by \cite{2015MNRAS.448.3026W}. Other estimates have been reported by \cite{2015ApJ...809...53C}, that predicted the number of joint EM and GW detections for different configurations of Advanced Virgo and Advanced LIGO and with \emph{Fermi}-GBM, considering the case of EM triggered GW searches. They found the  ranges 0.003-0.1 and 0.07-1 for the 2016-2017 and the design configurations respectively, that are consistent with our estimates. 

When GW searches triggered by \emph{Fermi}-GBM observations are included, the total number of GW detections almost double with respect to all-sky (and all-time) GW searches, for both the 2016-2017 and the design configuration; the same result is obtained for the expected number of joint EM and GW detections (see Fig. \ref{fig:SNR}).

\subsubsection{GRB afterglow emission}\label{sec:EMresults-afterglow}

In Tables \ref{tab:EMLAT1} and \ref{tab:EMLAT2} are shown the rates of GRB high-energy afterglows detectable by \emph{Fermi}-LAT, as well as the rates of events detectable both in EM and GW, for different values of the integration time $t_{\rm f}$ and different configurations of the interferometers, for a GW-alert latency of 0 s and 10 minutes respectively. 

\begin{table}[tbp]  
\begin{center}
\scalebox{0.88}{
\begin{tabular}{lllcccccc}
\hline \hline   \vspace{0.1mm}\\
 Integration &$E_\gamma$& EM  & EM and GW &  EM and GW \vspace{0.1mm}\\
Time  & & &  2016-2017  &  design   \vspace{0.1mm}\\
(s) & (ergs) & (yr$^{-1}$)&    (yr$^{-1}$)&  (yr$^{-1}$)  \vspace{0.2mm}\\
\hline \hline  
10 & 3.5$\times 10^{52}$& 0.5 (0.02 - 6.6) & 0.002 ($< 10^{-3}$ - 0.04) & 0.06 (0002 - 0.9) \vspace{0.3mm}\\
 & 1$\times 10^{49}$& 0.08 (0.002 - 1.1) & 0.002 ($< 10^{-3}$ - 0.04) &  0.05 ($< 10^{-3}$ - 0.6) \vspace{0.4mm}\\
10$^2$ & 3.5$\times 10^{52}$& 0.5 (0.02 - 6.6)& 0.002 ($< 10^{-3}$ - 0.04)  & 0.06 (0002 - 0.9) \vspace{0.4mm}\\
 & 1$\times 10^{49}$& 0.09  (0.002 - 1.2)& 0.002 ($< 10^{-3}$ - 0.04) &  0.05 ($< 10^{-3}$ - 0.6)\vspace{0.4mm}\\
10$^{3}$ & 3.5$\times 10^{52}$& 0.5 (0.02 - 6.6)& 0.002 ($< 10^{-3}$ - 0.04)  & 0.06 (0002 - 0.9)\vspace{0.3mm}\\
 & 1$\times 10^{49}$& 0.1  (0.002 - 1.2)& 0.002 ($< 10^{-3}$ - 0.04)  &  0.05 ($< 10^{-3}$ - 0.6)\vspace{0.4mm}\\
\hline\hline
\end{tabular}
}
\caption{\label{tab:EMLAT1} Expected rates of EM and GW detections for the 2016-2017 and the 2019+ (design) configurations, considering a 80 $\%$ duty cycle of the interferometers and a latency of 0 s; two values of $E_\gamma$ have been considered (see sec. \ref{sec:afterglow}). The reported estimates refer to the Standard model B, while the range of rates have been estimated considering the range of BNS merger rates reported by \protect\cite{2012ApJ...759...52D} (see Sec. \ref{sec:gw}). The rate of GW detections for the 2016-2017 configuration has been re-scaled to a 6-months observation period.} 
\end{center}
\end{table}

\begin{table}[tbp]  
\begin{center}
\scalebox{0.88}{
\begin{tabular}{lllcccccc}
\hline \hline   \vspace{0.1mm}\\
 Integration &$E_\gamma$& EM  & EM and GW &  EM and GW \vspace{0.1mm}\\
Time  & & &  2016-2017  &  design   \vspace{0.1mm}\\
(s) & (ergs) & (yr$^{-1}$)&    (yr$^{-1}$)&  (yr$^{-1}$)  \vspace{0.2mm}\\
\hline \hline  
10 & 3.5$\times 10^{52}$& 0.01 ($< 10^{-3}$ - 0.2) & 0.001 ($< 10^{-3}$ - 0.02) & 0.007 ($< 10^{-3}$ - 0.1) \vspace{0.3mm}\\
 & 1$\times 10^{49}$& $< 10^{-3}$ ($< 10^{-3}$ - $< 10^{-3}$)& $< 10^{-3}$ ($< 10^{-3}$ - $< 10^{-3}$) & $< 10^{-3}$ ($< 10^{-3}$ - $< 10^{-3}$) \vspace{0.4mm}\\
10$^2$ & 3.5$\times 10^{52}$& 0.3 (0.01 - 4.1)&  0.002 ($< 10^{-3}$ - 0.04) & 0.06 (0.002 - 0.9) \vspace{0.4mm}\\
 & 1$\times 10^{49}$& $< 10^{-3}$ ($< 10^{-3}$ - $< 10^{-3}$) & $< 10^{-3}$ ($< 10^{-3}$ - $< 10^{-3}$) & $< 10^{-3}$ ($< 10^{-3}$ - $< 10^{-3}$) \vspace{0.4mm}\\
10$^{3}$ & 3.5$\times 10^{52}$& 0.5 (0.02 - 6.6) & 0.002 ($< 10^{-3}$ - 0.04)  & 0.06 (0.002 - 0.9) \vspace{0.3mm}\\
 & 1$\times 10^{49}$& $< 10^{-3}$ ($< 10^{-3}$ - $< 10^{-3}$) & $< 10^{-3}$  ($< 10^{-3}$ - $< 10^{-3}$ ) & $< 10^{-3}$ ($< 10^{-3}$ - $< 10^{-3}$) \vspace{0.4mm}\\
\hline\hline
\end{tabular}
}
\caption{\label{tab:EMLAT2} Same as in Table 3, for a latency of 600 s.}
\end{center}
\end{table}

It can be seen that, for a latency of 0 s, there will be some chance to detect both the highest energetic GRBs ($E_\gamma=3.5 \times 10^{53}$ ergs) and the less energetic sources ($E_\gamma=1 \times 10^{49}$ ergs), with EM detection rates in the ranges (0.02 - 6.6) yr$^{-1}$ and (0.002 - 1.2) yr$^{-1}$ respectively. It can also be noted that this rate is almost independent on the integration time: all the simulated on-axis sources are located at lower distances with respect to GRB 090510, so their flux is intense enough to be detected with a short observing time. The rate of joint EM and GW detections is $<$ 1 when considering the 2016-2017 configuration of Advanced Virgo and Advanced LIGO, but when the interferometers will reach their final design sensitivity there will be some chance of a coincident EM and GW detection (the maximum rate of joint EM and GW detection is $\sim$ 1 yr$^{-1}$).
When a 600 s latency is considered (EM follow-up) it can be seen that, for the highest energetic GRBs, an integration time of $10^3$ s is needed to reach the same EM detection rate obtained for a 0 s latency; also in this case high sensitivity interferometers are needed for joint EM and GW detections. However, when the less energetic GRBs are considered, the rate of EM and joint EM and GW detections are both $< 1$.


It is important to recall that these results are based on the assumptions that all the short GRBs present a GeV extended emission, so they should be considered as upper limits.

\subsection{EM and GW detections - estimates based on the local short GRB rate}\label{sec:EMresults-localRate}
The rates of BNS mergers, EM (GRB prompt emission) and GW detections estimated from the local short GRB rate are reported in Table \ref{tab:EM-GBM2}, for different values of $\theta_j$. It can be seen that, for the fiducial value $\theta_j$=10$^\circ$, the rate of GW detections and of EM detections obtained with the value of ${\rm \rho_{sGRB}}$ by \cite{2015MNRAS.448.3026W} and \cite{2016arXiv160707875G}, model ``c'' are consistent, within the errors, with the ones previously obtained considering the Standard Model B by \cite{2012ApJ...759...52D} (see Tables \ref{tab:GW} and \ref{tab:EM-GBM}). 

\begin{table}[tbp]    
\begin{center}
\scalebox{0.88}{
\begin{tabular}{lllccccccc}
\hline \hline   \vspace{0.1mm}\\
$\theta_j$ & Model & BNS & EM & GW & GW \vspace{1mm}\\  
 & & & & \textbf{2016-2017} & \textbf{design}\vspace{1mm}\\ 
deg&  & yr$^{-1}$ & yr$^{-1}$ &  yr$^{-1}$ & yr$^{-1}$ \vspace{1mm}\\  
\hline
\vspace{1mm}\\  
 &  \cite{2015MNRAS.448.3026W} & 38200$^{+16810}_{-13000}$ & 0.2$^{+0.09}_{-0.07}$ & 67$^{+29}_{-23}$ & 1340$^{+590}_{-450}$ \vspace{2mm}\\
  0.3 & \cite{2016arXiv160707875G}, ``a''  & 7640$^{+1530}_{-2670}$ & 0.04$^{+0.008}_{-0.01}$& 13$^{+3}_{-5}$ &  267$^{+53}_{-94}$ \vspace{2mm}\\
&  \cite{2016arXiv160707875G}, ``c'' & 30560$^{+11460}_{-5730}$ & 0.16$^{+0.06}_{-0.03}$ & 53$^{+20}_{-10}$ & 1079$^{+401}_{-200}$\vspace{2mm}\\
 \hline \vspace{0.1mm}\\
 
 &  \cite{2015MNRAS.448.3026W} & 35$^{+15}_{-12}$ & 0.2$^{+0.09}_{-0.07}$ & 0.06$^{+0.03}_{-0.02}$ & 1.2$^{+0.5}_{-0.4}$\vspace{2mm}\\
 10 & \cite{2016arXiv160707875G}, ``a''  & 6.9$^{+1.4}_{-2.4}$ &  0.04$^{+0.008}_{-0.01}$ &  0.01$^{+0.002}_{-0.004}$ & 0.2$^{+0.05}_{-0.08}$ \vspace{2mm}\\
&  \cite{2016arXiv160707875G}, ``c'' & 28$^{+10}_{-5}$ &  0.16$^{+0.06}_{-0.03}$ & 0.05$^{+0.02}_{-0.01}$ & 1.0$^{+0.4}_{-0.2}$ \vspace{2mm}\\
  \hline \vspace{0.1mm}\\
  
 &  \cite{2015MNRAS.448.3026W} & 4$^{+1.7}_{-1.3}$ &0.2$^{+0.09}_{-0.07}$ & 0.005$^{+0.003}_{-0.002}$ & 0.14$^{+0.06}_{-0.05}$\vspace{2mm}\\
 30 & \cite{2016arXiv160707875G}, ``a''  & 0.8$^{+0.2}_{-0.3}$ &  0.04$^{+0.008}_{-0.01}$ & 0.001$^{+0.0003}_{-0.0005}$& 0.03$^{+0.005}_{-0.01}$\vspace{2mm}\\
&  \cite{2016arXiv160707875G}, ``c'' & 3$^{+1}_{-0.6}$ &  0.16$^{+0.06}_{-0.03}$ & 0.005$^{+0.002}_{-0.001}$& 0.1$^{+0.04}_{-0.02}$\vspace{2mm}\\
\hline\hline
\end{tabular}
}
\caption{\label{tab:EM-GBM2} Expected rates of BNS mergers, EM and GW detections, obtained using the values of $\rho_{\rm sGRB}$ reported in \cite{2015MNRAS.448.3026W} and \cite{2016arXiv160707875G}, models ``a'' and ``c''. The rate of GW detections for the 2016-2017 configuration has been re-scaled to a 6-months observation period. The reported uncertainties take into account the uncertainties in the estimates of $\rho_{\rm sGRB}$ (see \cite{2015MNRAS.448.3026W,2016arXiv160707875G}).} 
\end{center}
\end{table}

\begin{figure}
\begin{center}
\includegraphics[scale=1.]{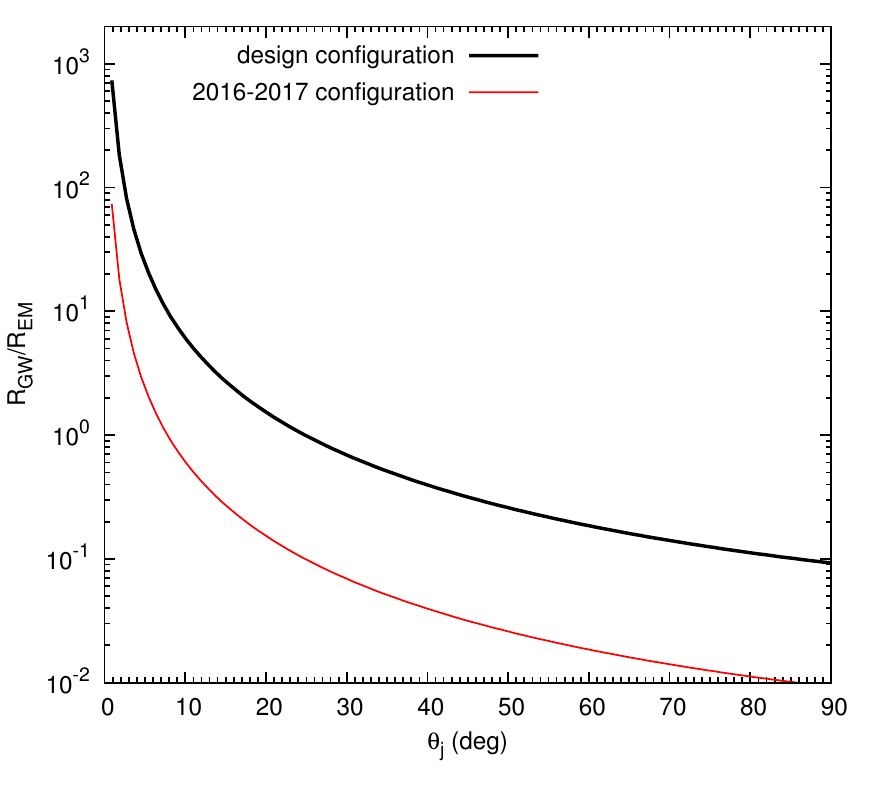}
\caption{Ratio between the rate of GW detections and the rate of EM detections with \emph{Fermi}/GBM,  as a function of $\theta_j$. The black (red) line refers to the design (2016-2017) configuration of the interferometers}\label{fig:ratio}
\end{center}
\end{figure}

We also estimated the ratio between the rate of GW detections and of EM detections ($R_{GW}/{\rm R_{EM}}$). This ratio only depends on $\theta_j$ and on the characteristics of the GW and EM detectors; it does not depend on $\rho_{\rm sGRB}$, since both $R_{GW}$ and ${\rm R_{EM}}$ are proportional to $\rho_{\rm sGRB}$. The results are shown in fig. \ref{fig:ratio}. It can be seen that, as expected, $R_{GW}/{\rm R_{EM}}$ decreases by increasing $\theta_j$. For $\theta_j$=10$^\circ$, the ratio is $\sim$ 6 (0.6) when considering the design (2016-2017) configuration of Advanced Virgo and Advanced LIGO. The comparison of these predictions with future joint GW and EM observation would provide a constraint on the jet opening angle of GRBs and/or on the fraction of short GRBs having a BNS progenitor.

\section{Conclusions}\label{sec:concl}

We presented a detailed study on the expectations for joint GW and high-energy EM observations of binary mergers with the interferometers Advanced Virgo and Advanced LIGO and with the $\gamma$-ray observatory \emph{Fermi}, with focus on BNS systems. This study uses: i) the construction of a sample of BNS merging systems populating the local universe; ii) the simulation of the associated GW signals and the estimate of their detectability with Advanced Virgo and Advanced LIGO considering their future evolving configurations and iii) the simulation of the associated high-energy EM signals (short GRBs) and the assessment of their detectability with \emph{Fermi}. Our approach differentiates from previous works on joint EM and GW observations in several points. First of all, in this work we investigated joint EM and GW observations combining accurate population synthesis modeling (or accurate estimates of the local short GRB rates) with pipelines specifically developed to provide low-latency GW sky localization (i.e. BAYESTAR), needed to perform the EM follow-up of GW events; furthermore, we focused on the high-energy emission, in particular on $\gamma$-rays and included in our study both the GeV prompt and afterglow emission of short GRBs.

When using population synthesis models to estimate the BNS merger rate, we have found that the expected number of GW detections is in the range 0.001 - 0.7 and 0.04 - 15 for a six-month science run with the 2016-2017 configuration and for a 1-year science run with the design configuration respectively; the ``reference'' values are 0.05 and 1  respectively: these values are consistent with the estimates based on the local short GRB rate if a jet opening angle of 10$^\circ$ is assumed. The typical GW sky localization of these events is from hundreds to thousands of square degrees: this underlines the importance of having large FOV telescopes such as \emph{Fermi} to perform the EM follow-up of GW events. These results are consistent with previous estimates reported in literature.

We have also shown that the expected rate of coincident GRB prompt emission and GW signal detections is low during the GW observing run planned for 2016-2017; however, as the interferometers approach their final design sensitivity, this rate will increase and there could be up to $\sim$ 3 joint detections for a 1-year observing run. The rates are expected to increase when high-energy EM triggers are provided: in this case the GW search threshold can be lowered by $\sim$ 17 \% with respect to all sky, all time searches. This will almost double the number of GW detections and of joint EM and GW detections.
 
When focusing on the GRB afterglow emission we found that, in the case of no latency between the EM and GW observations, there is some chance of a joint EM and GW detection when Advanced Virgo and Advanced LIGO will reach their design sensitivity, with an expected rate in the range (0.002 - $\sim$ 1) yr$^{-1}$. When a latency of 600 s is considered, for the highest energetic GRBs an integration time of $\sim$ 10$^{3}$ s is needed to reach the same detection rates; when the less energetic GRBs are considered ($E_\gamma=10^{49}$ erg), the rate of joint EM and GW detections is $<$ 1 yr$^{-1}$.

The results here presented have shown that the EM observations at high energies with \emph{Fermi} represent a promising instrument to identify the EM counterpart of GW transient events detected by Advanced Virgo and Advanced LIGO (see also \cite{2016ApJ...826L...6C,2016ApJ...823L...2A}). Owing to its large FOV and the sharp PSF of LAT, \emph{Fermi} could observe the prompt $\gamma$-ray emission of short GRBs associated with GW transients and provide refined sky localization to other telescopes: this will allow the EM follow-up of the GW event by other instruments having smaller FOV and covering different energy ranges, allowing the identification and the multi-wavelength characterization also of the fainter sources. This work represents an important step describing the potential of joint GW and EM observations. A comparison of the future observations with the joint GW and EM detection rates here estimated could help to shed light on the physics of compact objects and will allow to put constraints on the association between short GRBs and BNS systems. Furthermore, this work could be helpful in defining the best EM follow-up strategies for the future observation runs of Advanced Virgo and Advanced LIGO.

In order to make the results of this work accessible to  the community, we will soon provide an online database, that will be available at \url{virgopisa.df.unipi.it/HEGWFollowup}.

\acknowledgments
We thank the anonymous referee for the useful comments, that helped us to improve the paper. We also thank Gang Wang for his useful suggestions. 
BP, MR and MB have been supported by the contract \mbox{FIRB-2012-RBFR12PM1F} of the Ministry of Education, University and Research (MIUR).



\bibliographystyle{JHEP}
\bibliography{EMFup}



\end{document}